\documentclass[a4paper,twoside]{article}

\usepackage{epsfig}
\usepackage{subcaption}
\usepackage{calc}
\usepackage{graphicx}
\usepackage{booktabs}
\usepackage{amssymb}
\usepackage{amstext}
\usepackage{amsmath}
\usepackage{amsthm}
\usepackage{mathtools}
\usepackage{multicol}
\usepackage{pslatex}
\usepackage{apalike}
\usepackage{algorithm2e}
\usepackage{xcolor}
\usepackage[bottom]{footmisc}
\usepackage{SCITEPRESS}

\renewcommand{\orcidAuthor}[1]{}

\begin{document}

\title{Catalogue-Grounded Multimodal Attribution for Museum Video under Resource and Regulatory Constraints}

\author{\authorname{Minsak Nanang\sup{1}, Adrian Hilton\sup{1} and Armin Mustafa\sup{1}}
\affiliation{\sup{1}University of Surrey, England, United Kingdom}
\affiliation{\sup{1}Faculty of Engineering and Physical Sciences}
\email{jn00767@surrey.ac.uk, armin.mustafa@surrey.ac.uk}
}


\abstract{Museum audiovisual archives are important records of visual culture, yet in-gallery video remains difficult to search because it lacks catalogue-compatible metadata. We propose CaGMulA, catalogue-grounded multimodal attribution for museum video: a locally deployable pipeline that combines video-language models, structured catalogue retrieval, and abstention-aware decision policies. The system summarises visible artworks, generates catalogue-style descriptions, and attributes title and artist only when evidence is sufficient. We evaluate VideoLLaMA2 and Qwen2-VL across four model configurations and four attribution modes on 18 in-gallery videos. Open-ended generation fails under conservative conditions, producing 0\% coverage across models. By contrast, closed-set catalogue retrieval reaches 100\% coverage with 62\% accepted title--artist correctness, while a verifier-gated hybrid mode reduces coverage to 50\% but improves accepted correctness to 78\%. The results show that reliable museum-video attribution is driven less by unrestricted generation than by catalogue grounding, visual retrieval, and selective abstention for curatorial review.}

\onecolumn \maketitle \normalsize \setcounter{footnote}{0} \vfill

\section{\uppercase{Introduction}}

Museum and gallery audiovisual archives are important records of visual culture, yet their moving-image content remains difficult to search, reuse, and analyse at scale. Unlike digitised still-image collections, in-gallery video often contains partial artwork views, wall labels, visitors, architectural context, camera motion, and uneven visibility. These characteristics make museum audiovisual material a challenging setting for computer vision: the system must reason over visual evidence, generate useful descriptions, and connect uncertain observations to structured cultural records. We therefore frame museum video metadata generation as \textbf{catalogue-grounded multimodal attribution} for cultural heritage collections.

This setting is shaped by two institutional constraints. First, museum audiovisual content is often rights-restricted or operationally sensitive, making local deployment and data sovereignty important. Second, attribution errors are asymmetric: an incorrect title or artist can propagate into search, scholarship, collections management, and internal decision-making. A practical system should therefore prefer conservative attribution with explicit abstention when the available evidence is weak~\cite{aiarchives_review}. Prior work on audiovisual cultural heritage similarly identifies archive disclosure, precise metadata, legal constraints, and usable access as central challenges for making moving-image collections discoverable and reusable~\cite{hauttekeete2011}.

Multimodal large language models (MLLMs) provide a possible route toward this automation. Systems such as LLaVA~\cite{llava}, Qwen-VL~\cite{qwenvl}, BLIP-2~\cite{blip2}, Flamingo~\cite{flamingo}, Video-LLaMA~\cite{videollama}, and VideoLLaMA2~\cite{videollama2} can generate rich language from visual inputs. Museum attribution, however, requires more than open-ended generation. A system must ground uncertain visual observations in a finite catalogue, produce catalogue-compatible metadata, and abstain when the evidence is insufficient.

We make the following contributions:

\begin{itemize}
    \item \textbf{A retrieval-first formulation for museum moving-image attribution.}
    We formulate artwork identification in museum video as catalogue-grounded attribution rather than open-ended visual question answering, aligning the task with finite, curated, and institutionally meaningful collections.

    \item \textbf{A multimodal pipeline for catalogue-compatible metadata.}
    We introduce a model-agnostic pipeline combining video-language generation, wall-label transcription, structured catalogue matching, visual catalogue retrieval, and conservative abstention. The system produces descriptive metadata while treating title--artist attribution as a selective decision that may return \texttt{not visible}.

    \item \textbf{A pluggable VLM backend for controlled comparison.}
    We evaluate VideoLLaMA2 and Qwen2-VL under identical prompts, retrieval indices, catalogue matching, and decision policies, helping isolate the effects of model family, fine-tuning, retrieval, and verifier-gated abstention.

    \item \textbf{An ablation across four attribution modes.}
    We compare label-first prompting, free generation, retriever-only attribution, and hybrid verification across four model configurations on 18 in-gallery videos. The results show that free generation fails under conservative conditions, catalogue retrieval provides the strongest high-coverage baseline, and verifier-gated abstention improves accepted precision by rejecting uncertain cases.
\end{itemize}

\paragraph{Why catalogue grounding rather than classification.}

We do not formulate attribution as a fixed $N$-way classification problem because museum catalogues are not static. Collections change as works are acquired, lent, reattributed, re-catalogued, or deaccessioned. A catalogue-grounded architecture can update its retrieval index without retraining the underlying multimodal model.

This formulation also supports the broader metadata-generation problem. Generative backbones remain useful for descriptions, genres, scene summaries, and multi-artwork observations, while identity attribution is bounded by retrieval, structured catalogue records, and explicit abstention.


\section{\uppercase{Related Work}}

\subsection{Multimodal Large Language Models}

Multimodal large language models combine visual encoders with large language model backbones to support image- and video-conditioned generation, instruction following, and visual reasoning. Representative systems include Flamingo~\cite{flamingo}, BLIP-2~\cite{blip2}, LLaVA~\cite{llava}, Qwen-VL~\cite{qwenvl}, and video-oriented extensions such as Video-LLaMA~\cite{videollama} and VideoLLaMA2~\cite{videollama2}.

Building on these advances, our work studies how MLLMs can be adapted from general video understanding to catalogue-grounded cultural heritage attribution. The focus is on producing \emph{catalogue-compatible fields}, linking uncertain visual evidence to structured museum records, and enforcing conservative attribution under institutional constraints. To examine whether the proposed pipeline generalises beyond one model family, we evaluate two architecturally distinct backends: VideoLLaMA2, which combines a SigLIP vision encoder with a Qwen2 language backbone, and Qwen2-VL~\cite{qwen2vl}, which uses its native vision-language architecture with dynamic-resolution visual processing.

\subsection{Grounding and Retrieval-Augmented Generation}

Grounding in language models is commonly implemented through retrieval-augmented generation (RAG)~\cite{rag} or systems that integrate retrieval directly into model computation~\cite{retllm}. Related approaches incorporate structured knowledge sources~\cite{kglm} or teach models to invoke external tools~\cite{toolformer}.

Our retrieval target differs from open-domain RAG because it is a \emph{finite, structured museum catalogue}. Textual matching is performed over canonical catalogue fields using lightweight IDF-weighted Jaccard similarity, character-level matching, and alias expansion, drawing conceptually from entity-linking and record-linkage approaches~\cite{entitylinking,recordlinkage}. We additionally evaluate visual catalogue retrieval, in which video frames or artwork-focused crops are matched directly against catalogue images before a generative model is permitted to commit to an attribution. This formulation enables explicit acceptance thresholds, closed-set candidate generation, and auditable failure modes.

\subsection{Selective Prediction and Abstention}

Selective prediction studies when models should defer or abstain~\cite{geifman2017selective,abstention}. This is particularly relevant to museum attribution because false-positive identity assignments can have substantial downstream consequences.

We implement abstention through a configurable text-matching framework (Section~\ref{sec:abstention-system}) and a conservative retrieval--verifier policy. The system accepts high-confidence catalogue matches when sufficient evidence is present and abstains when retrieval confidence, matching margins, or verification evidence are insufficient. This design allows operators to control the precision--coverage trade-off according to institutional risk tolerance.

\subsection{AI for Cultural Heritage}

Machine learning for cultural heritage has addressed style analysis, art-historical pattern discovery, and content-based retrieval~\cite{elgammal2018shaped,mensink2014,garcia2019context}. Our work differs in three respects. First, we target \emph{in-gallery video} rather than static artwork images. Second, we prioritise locally deployable architectures suitable for rights-sensitive collections. Third, we treat catalogue grounding with explicit abstention as the central mechanism for safe identity attribution.


\section{\uppercase{Methodology}}
\label{sec:methodology}

Our goal is to transform in-gallery video footage captured while moving through gallery rooms into searchable, catalogue-compatible metadata. Target fields include title, artist, subject, genre, and textual description. This section describes the complete pipeline, the abstention mechanism, and the pluggable model architecture.

\subsection{System Overview}
\label{sec:system-overview}

The pipeline receives an in-gallery video $v$ and a structured painting catalogue $\mathcal{C}$, and produces: (i) a multi-artwork summary listing up to three artworks with locations and visual descriptions; (ii) a catalogue-style description and genre label for the main work; and (iii) catalogue-grounded identity fields, or explicit \texttt{not visible} abstention.

\begin{figure*}[t]
    \centering
    \includegraphics[width=\textwidth]{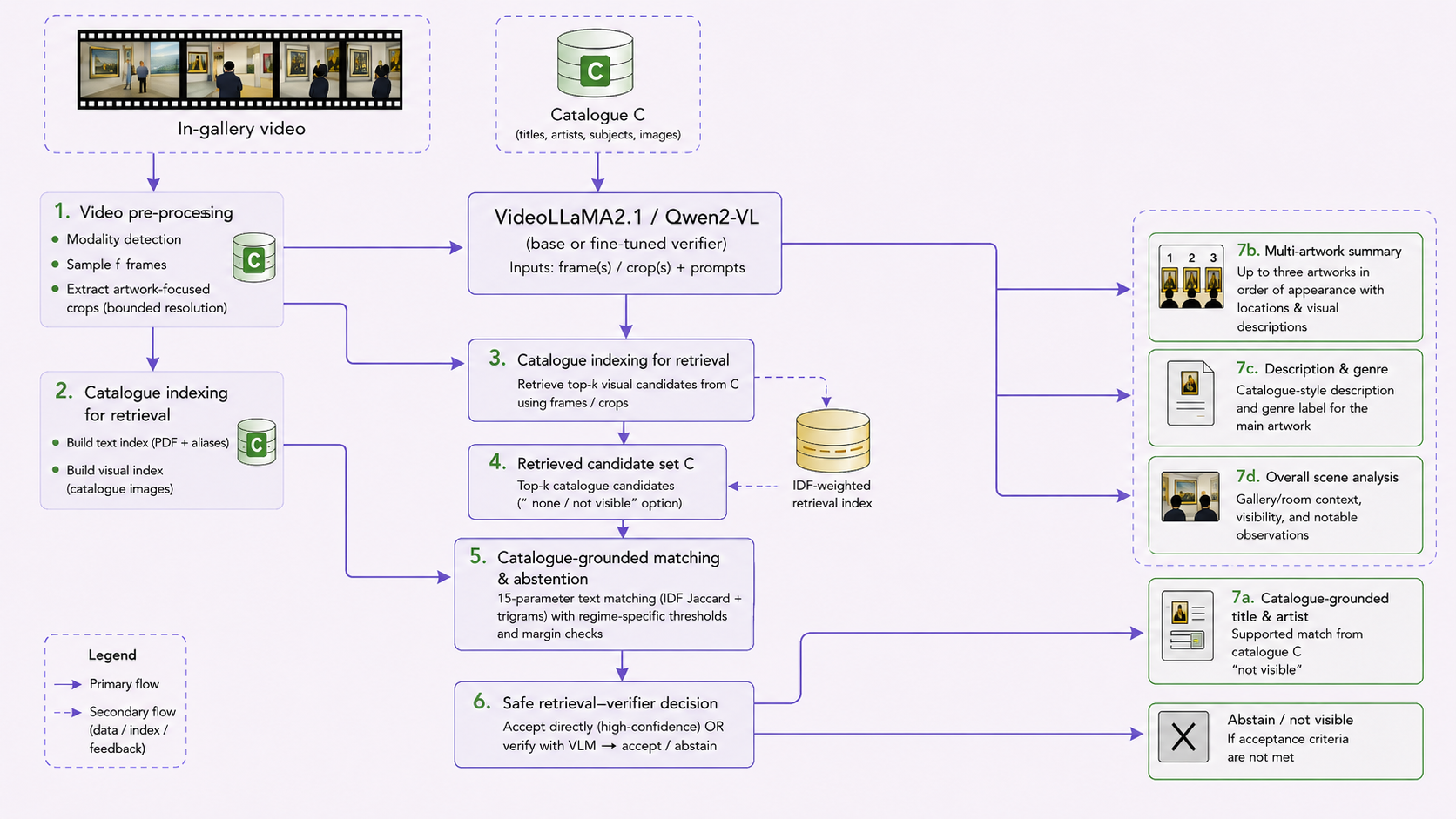}
    \caption{High-level overview of the catalogue-grounded multimodal attribution pipeline. Video evidence is converted into artwork-focused visual inputs, matched against a finite museum catalogue, verified when necessary, and passed through an abstention-aware decision layer before metadata is returned.}
\label{fig:gc-pipeline}
\end{figure*}

The pipeline proceeds in seven stages, with identity attribution spanning Stages~3--6 and descriptive metadata produced in Stage~7.

\paragraph{1. Media preprocessing.}
We detect modality, sample $T$ frames from each in-gallery video, and extract artwork-focused crops at bounded resolution. These frames or crops form the visual evidence supplied to the backend model and to the visual retrieval index.

\paragraph{2. Catalogue indexing for retrieval.}
From the catalogue JSON we construct a finite index $\mathcal{C}$ containing titles, artists, subjects, aliases, and catalogue images. Text fields are normalised, parenthetical and quoted title variants are retained as aliases, domain-specific stopwords are removed, and IDF weights are computed over catalogue entries. We also build a visual index from catalogue images so that video frames and crops can be matched directly to known works.

\paragraph{3. Visual candidate retrieval.}
Artwork-focused frames or crops are retrieved against the catalogue image index to obtain top-$k$ visual candidates from $\mathcal{C}$. This step makes attribution closed-set and catalogue-grounded before any model is allowed to commit to an identity.

\paragraph{4. Retrieved candidate set.}
The retrieved set is converted into a candidate-verification prompt containing plausible catalogue identities and a \texttt{none/not visible} option. This constrains the task from open-ended naming to evidence-based candidate selection.

\paragraph{5. Catalogue-grounded matching and abstention.}
Model outputs and catalogue candidates are scored using IDF-weighted token similarity and character-trigram matching with regime-specific thresholds and margin checks (Section~\ref{sec:abstention-system}). This produces an auditable accept/abstain decision record.

\paragraph{6. Safe retrieval--verifier decision.}
High-confidence retrieval matches may be accepted directly when score and margin thresholds are met. Otherwise, the verifier is consulted; agreement can support acceptance, while disagreement or insufficient evidence triggers abstention rather than allowing an unsupported override.

\paragraph{7. Metadata outputs.}
The system returns (7a) catalogue-grounded title and artist or \texttt{not visible}, (7b) a multi-artwork summary, (7c) a catalogue-style description and genre label, and (7d) an overall scene analysis. These outputs remain useful even when identity attribution abstains.

\subsection{Configurable Abstention System}
\label{sec:abstention-system}

A core technical contribution is the abstention system that governs when the pipeline commits to a catalogue match. Rather than a single threshold, we implement a structured decision framework with 15 tunable parameters organised into three operating regimes.

\paragraph{Signal collection.}
The pipeline gathers three signals from the model: a title guess, an artist guess, and a subject description. Under \emph{strict abstention mode}, uncertain outputs such as ``not sure'', ``unknown'', or ``I don't know'' are dropped entirely; under \emph{relaxed mode}, they are passed to the matcher as-is.

\paragraph{Regime selection.}
Based on which signals are available and their strength, the matcher selects one of three regimes:

\begin{enumerate}
    \item \textbf{Artist-driven}: activated when the best artist similarity score exceeds $\tau_{\mathrm{artist}}$ (default 0.45). Uses weights $(w_a,w_t,w_s)=(0.46,0.36,0.18)$.
    \item \textbf{Title-driven}: the default regime when a title guess is available. Uses weights $(w_t,w_s)=(0.78,0.22)$.
    \item \textbf{No-title fallback}: when only artist and subject are available. Uses weights $(w_a,w_s)=(0.70,0.30)$.
\end{enumerate}

\paragraph{Scoring.}
Each catalogue entry is scored using a weighted combination of token-level IDF-Jaccard similarity and character-trigram Jaccard similarity:

\begin{equation}
S_{\mathrm{alias}}(g,c)
=
\alpha J_{\mathrm{IDF}}(g,c)
+
(1-\alpha)J_{\mathrm{tri}}(g,c),
\label{eq:alias-score}
\end{equation}

where $\alpha$ (default 0.65) controls the balance. For entries with multiple title aliases, we take the maximum score across aliases.

\paragraph{Margin-based acceptance.}
The margin is computed as the score difference between the best candidate and the first \emph{distinct} candidate. This prevents duplicate catalogue entries from collapsing the margin to zero. Acceptance requires \emph{both} the combined score and the margin to exceed regime-specific thresholds. For example, in the title-driven regime:

\begin{equation}
\text{accept if }
S_{\mathrm{title}} \geq \tau_t
\quad \land \quad
\Delta \geq \mu_t,
\label{eq:title-acceptance}
\end{equation}

where $\tau_t=0.52$ and $\mu_t=0.05$ by default, or if the combined score satisfies $S_{\mathrm{comb}}\geq\tau_c=0.44$ with margin $\Delta\geq\mu_c=0.04$.

\begin{table*}[t]
\caption{Text-matching abstention configuration.}
\label{tab:abstention-params}
\centering
\small
\begin{tabular}{@{}p{0.18\textwidth}p{0.50\textwidth}p{0.16\textwidth}@{}}
\toprule
\textbf{Parameter} & \textbf{Description} & \textbf{Default} \\
\midrule
\multicolumn{3}{@{}l}{\emph{Artist-driven regime}} \\
$\tau_{\mathrm{artist}}$ & Artist score to activate regime & 0.45 \\
$\tau_{\mathrm{artist,accept}}$ & Combined score for acceptance & 0.38 \\
\midrule
\multicolumn{3}{@{}l}{\emph{Title-driven regime}} \\
$\tau_t$ & Direct title score accept & 0.52 \\
$\mu_t$ & Title margin required & 0.05 \\
$\tau_c$ & Combined score accept & 0.44 \\
$\mu_c$ & Combined margin required & 0.04 \\
\midrule
\multicolumn{3}{@{}l}{\emph{Fallback regime}} \\
$\tau_f$ & Fallback accept threshold & 0.42 \\
$\mu_f$ & Fallback margin & 0.04 \\
\midrule
\multicolumn{3}{@{}l}{\emph{Scoring weights}} \\
$\alpha$ & Token vs. trigram weight & 0.65 \\
$(w_a,w_t,w_s)$ & Artist-strong regime weights & $(0.46,0.36,0.18)$ \\
$(w_t,w_s)$ & Title-driven regime weights & $(0.78,0.22)$ \\
$(w_a,w_s)$ & No-title fallback weights & $(0.70,0.30)$ \\
\midrule
\multicolumn{3}{@{}l}{\emph{Pipeline behaviour}} \\
\texttt{label\_first} & Try label transcription first & True \\
\texttt{strict\_abstention} & Drop uncertain model outputs & True \\
\texttt{force\_visual} & Also run visual Q\&A if label read & False \\
\bottomrule
\end{tabular}
\end{table*}

\paragraph{Safe retrieval--verifier policy.}
For retrieval-grounded attribution, we introduce a second abstention layer that operates over the top-$k$ visual catalogue candidates. The retriever-only mode accepts the top candidate when its score exceeds a retrieval threshold. The hybrid-verification mode applies a safer policy: accept the top-1 candidate directly only when it is high-confidence ($s_1\geq0.88$) and separated from the second candidate by a sufficient margin ($\Delta\geq0.04$); otherwise consult the verifier. If the verifier agrees with the top candidate and the top score exceeds a lower acceptance threshold ($s_1\geq0.75$), the system accepts. If the verifier disagrees, the default policy abstains rather than allowing the verifier to override a strong retrieval signal.

\paragraph{Abstention output.}
Each decision produces a structured record containing the regime used, the decision (accept/abstain), the reasoning, scores, margin, and which thresholds were applied. This provides full auditability for curatorial review.

\subsection{Pluggable Backend Architecture}
\label{sec:backend-arch}

To enable controlled comparison across model families, we introduce a \texttt{ModelBackend} abstraction that decouples the inference interface from the underlying vision-language model. Each backend implements three methods: \texttt{load(path, quantise)}, \texttt{preprocess(media, modality)}, and \texttt{generate(media, prompt, params)}. The pipeline, prompts, catalogue matching, retrieval, and abstention system are identical across backends, helping isolate the effect of model choice.

We implement two backends:

\begin{itemize}
    \item \textbf{VideoLLaMA2}: wraps the native \texttt{mm\_infer} interface with SigLIP vision tower, MLP projector, and Qwen2 decoder.
    \item \textbf{Qwen2-VL}: wraps \texttt{Qwen2VLForConditionalGeneration} with deterministic frame extraction and per-frame pixel budgets.
\end{itemize}

Both backends support 4-bit quantised loading via \texttt{BitsAndBytesConfig} for deployment on consumer GPUs with approximately 12\,GB VRAM.


\section{\uppercase{Data and Fine-Tuning}}

\subsection{Dialogue Construction from Catalogues}
\label{sec:dialogue-construction}

We construct catalogue-derived image dialogue pairs from a painting catalogue to teach the model museum-grade behaviour: concise identification when evidence is present, and explicit abstention when it is not. For each catalogue entry $c\in\mathcal{C}$, we synthesise a short dialogue mirroring a curator querying a multimodal assistant. Questions are sampled from paraphrased templates for each functional slot (title, artist, subject, description, and genre). Answers are populated from catalogue fields with lightweight normalisation to ensure identification targets are short and consistent.

\begin{figure}[t]
    \centering
    \includegraphics[width=\linewidth]{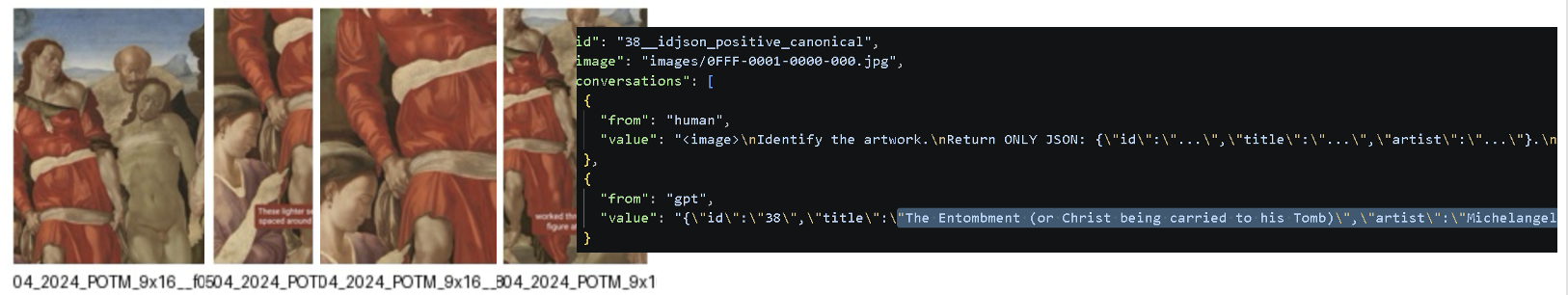}
    \caption{Construction of catalogue-derived multimodal training data. Structured museum catalogue records and associated artwork images are converted into catalogue-conditioned supervision for identification, candidate verification, hard-negative comparison, and explicit abstention.}
\label{fig:data-pipe}
\end{figure}

\paragraph{Synthetic abstention augmentation.}
With a small probability $p_{\mathrm{abs}}$, we synthesise abstention examples by prefixing the user question with a visibility cue and forcing the target to ``not visible''. When applied to a sample, auxiliary identification tasks are skipped to avoid contradictory supervision.

\paragraph{Alias construction.}
From $c.\texttt{title\_raw}$ we extract alternative title strings by taking the primary segment before `;' and harvesting parenthetical or quoted variants. These aliases support partial-match retrieval at inference but are not used as training targets.

\paragraph{Candidate-verification construction.}
To support retrieval-grounded attribution, we additionally construct candidate-choice examples. Each example pairs an image, frame, or crop with a short list of retrieved catalogue candidates and a ``none/not visible'' option. The target answer is a constrained JSON object selecting the correct candidate or abstaining. Hard negatives are sampled from visually similar catalogue entries, forcing the verifier to compare plausible alternatives rather than memorise frequent titles.

\subsection{Training Configuration}
\label{sec:training-config}

We apply LoRA~\cite{lora} fine-tuning to two model families using the same museum-specific supervision objective. Both fine-tuned models are trained as retrieval-conditioned verifiers rather than as unconstrained title generators. The training data comprise the same preserved atomic hybrid-verifier dataset containing 672 records. Each record pairs an artwork-focused crop with catalogue-conditioned supervision, including identification JSON, candidate verification, hard-negative comparison, and explicit abstention examples. This alignment ensures that both backbones learn the same deployment role: given visual crop evidence and catalogue candidates, either select a supported catalogue identity or return an abstention target when the evidence is insufficient.

\paragraph{VideoLLaMA2.1-7B-16F.}
VideoLLaMA2.1 uses a SigLIP vision tower, an MLP projector, and a Qwen2 language decoder. We fine-tune it with LoRA on the language projection layers while retaining the native VideoLLaMA2 visual front-end. The model is trained on the same 672 preserved atomic hybrid-verifier records used for Qwen2-VL, for 336 optimisation steps over two epochs. Training uses 4-bit NF4 base loading, LoRA rank $r=16$, $\alpha=32$, dropout 0.05, learning rate $5\times10^{-5}$, batch size 1, gradient accumulation 4, and \texttt{paged\_adamw\_8bit}. The final training loss is 0.8006. In the evaluation pipeline, the resulting merged model is used as a retrieval-conditioned verifier under the same safe decision policy as Qwen2-VL.

\paragraph{Qwen2-VL-7B-Instruct.}
Qwen2-VL uses its native vision encoder, vision merger, and Qwen2 language model. We fine-tune it on the same 672 preserved atomic hybrid-verifier records, also for 336 steps over two epochs. Training uses 4-bit loading, LoRA rank $r=16$, $\alpha=32$, dropout 0.05, learning rate $5\times10^{-5}$, batch size 1, gradient accumulation 4, bounded crop resolution, and LoRA on language attention projections with lightweight vision-merger adaptation. The final training loss is 1.1430. As with VideoLLaMA2, the merged model is evaluated as a retrieval-conditioned verifier and is not treated as a standalone free-form artwork identifier.

Training is therefore symmetric at the task level but not identical at the architectural level. Both fine-tuned models receive the same dataset, record count, optimisation horizon, LoRA rank, learning rate, batch schedule, and verifier objective. The difference lies in the multimodal front-end: VideoLLaMA2 encodes visual evidence through SigLIP and an MLP projector, whereas Qwen2-VL uses its native vision encoder and merger. We therefore interpret downstream differences as a combination of backbone architecture, visual encoding pathway, and calibration under the shared retrieval-abstention policy, rather than as a difference in training objective.

\begin{figure}[t]
    \centering
    \includegraphics[width=\linewidth]{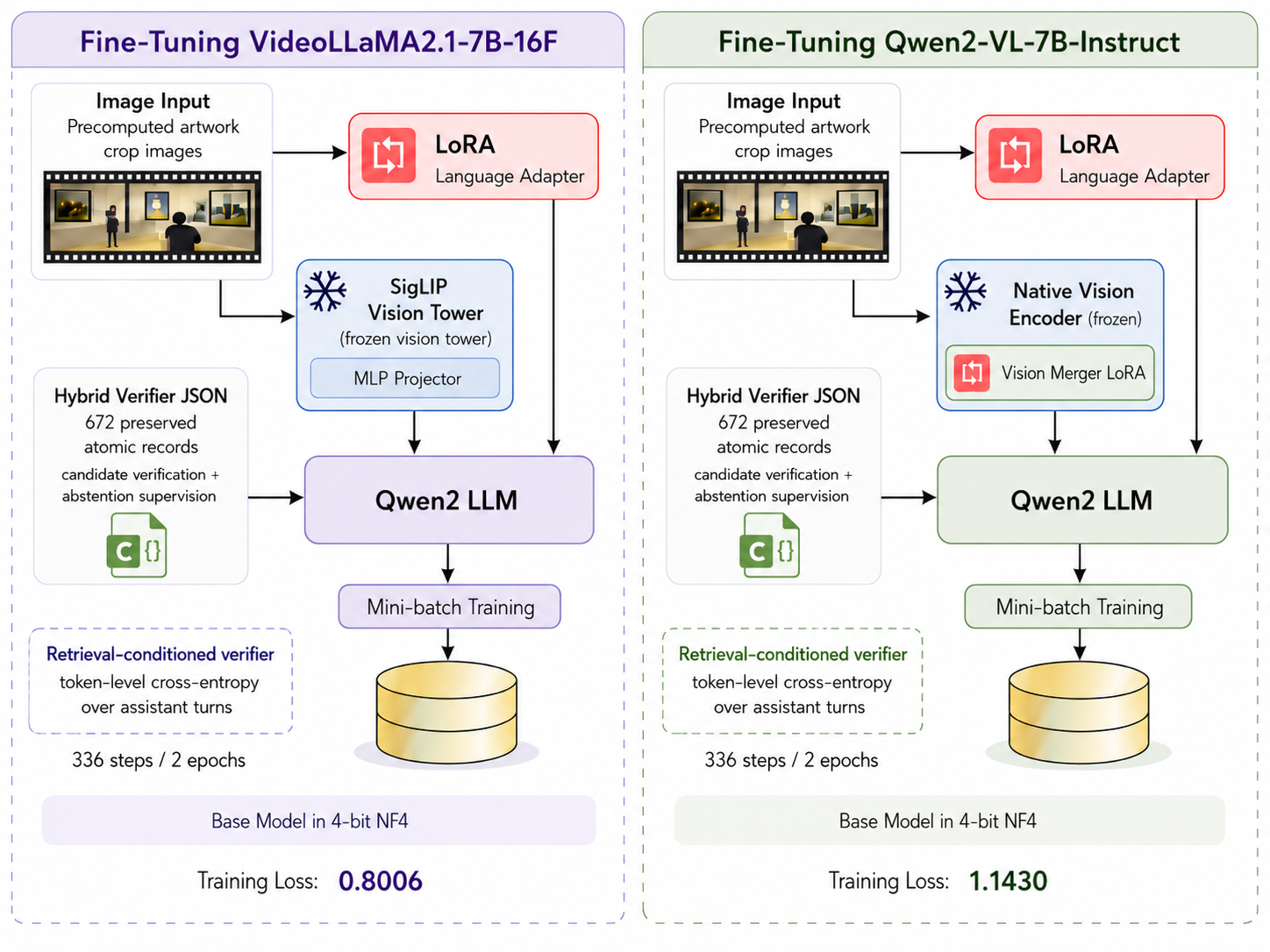}
    \caption{Model architecture and fine-tuning setup. Both VideoLLaMA2.1 and Qwen2-VL are fine-tuned as retrieval-conditioned verifiers on the same preserved atomic hybrid-verifier dataset. The supervision objective is shared, while the visual front-ends differ: VideoLLaMA2.1 uses a SigLIP vision tower and MLP projector, whereas Qwen2-VL uses its native vision encoder and vision merger.}
\label{fig:ft-models}
\end{figure}

\subsection{Training--Inference Format Alignment}
\label{sec:format-alignment}

A significant finding of our work is that \emph{training--inference format alignment} is critical for multimodal fine-tuning. The training protocol therefore aligns both fine-tuned backbones with the retrieval-verifier deployment path. Rather than training on generic catalogue dialogue and evaluating on candidate verification, both VideoLLaMA2.1 and Qwen2-VL are fine-tuned using the same crop-based hybrid-verifier records that appear at inference time. Precomputed artwork crops are treated as image inputs, and the textual target is constrained to catalogue-conditioned verification or abstention.

This change addresses a limitation of earlier Qwen2-VL experiments, where native video loading in an interactive setting and extracted-frame batch evaluation used different visual pathways. The current setup avoids this mismatch by using deterministic crop/image records for both training and evaluation. It also clarifies the role of fine-tuning: the model is not asked to hallucinate a title from open-ended visual memory, but to verify whether retrieved catalogue candidates are supported by the visible evidence. The safe hybrid policy remains essential, because even a fine-tuned verifier should not override a strong retrieval signal unless the acceptance thresholds and margins are satisfied.

\subsection{Training Objective}
\label{sec:training-objective}

We fine-tune by minimising token-level cross-entropy over assistant turns, using the standard instruction-tuning masking scheme in which user tokens are ignored. For target token sequence $y=(y_1,\ldots,y_L)$ conditioned on multimodal input $x$, the objective is

\begin{equation}
\mathcal{L}_{\mathrm{CE}}
=
-\sum_{t=1}^{L}
m_t
\log p_{\theta}
\left(
y_t \mid y_{<t},x
\right),
\label{eq:training-objective}
\end{equation}

where $m_t=1$ for supervised assistant tokens and $m_t=0$ for masked prompt or user tokens.

The deployment setting is compute-limited, so the training protocol is engineered for stability under low memory. The objective is shared across both fine-tuned backbones: learn catalogue-grounded verification over retrieved candidates while preserving explicit abstention behaviour. This objective supports the curatorial aim of high-precision metadata insertion: rich descriptions and genres may be generated by the broader pipeline, but title--artist attribution is treated as a selective decision grounded in catalogue evidence.


\section{\uppercase{Experiments and Results}}
\label{sec:experiments}

\subsection{Experimental Setup}

\paragraph{Evaluation data.}
We evaluate on 18 in-gallery videos from the partner institution's archive. Of these, 13 have ground-truth title and artist annotations, while the remaining videos are included for robustness observation and are not counted as correct/incorrect in the title--artist metrics. Videos include museum walkthrough footage with partial artwork views, camera motion, changing scale, label visibility, and visually similar catalogue alternatives.

\paragraph{Model configurations.}
We compare four model configurations, all evaluated under the same pipeline, prompts, catalogue, retrieval index, and abstention policy:

\begin{enumerate}
    \item \textbf{VL2-Base}: VideoLLaMA2.1-7B-16F zero-shot (no museum-specific training).
    \item \textbf{VL2-7B-FT-Merged}: VideoLLaMA2.1-7B-16F fine-tuned and merged as a retrieval-conditioned verifier.
    \item \textbf{Q2VL-7B-Base}: Qwen2-VL-7B-Instruct zero-shot.
    \item \textbf{Q2VL-7B-FT-Merged}: Qwen2-VL-7B-Instruct fine-tuned and merged as a retrieval-conditioned verifier.
\end{enumerate}

\paragraph{Attribution modes.}
Each configuration is evaluated under four modes:

\begin{enumerate}
    \item \textbf{label-first}: attempt wall-label transcription before visual attribution;
    \item \textbf{free-generation}: ask the VLM to identify the artwork directly without retrieval;
    \item \textbf{retriever-only}: accept the top visual catalogue retrieval candidate under thresholding;
    \item \textbf{hybrid-verify}: apply the safe retrieval-verifier policy with abstention.
\end{enumerate}

\paragraph{Metrics.}
For each GT-labelled video we report: (i) \emph{coverage}: rate of non-abstaining predictions; (ii) \emph{title accuracy}; (iii) \emph{artist accuracy}; (iv) \emph{both}: title and artist jointly correct; and (v) \emph{AcceptBoth}: title--artist correctness among accepted predictions only. The final metric is especially important in museum settings because it measures precision after abstention.

\subsection{Main Results}
\label{sec:main-results}

Table~\ref{tab:main-results} reports 288 evaluations across 16 configuration/mode groups. The strongest result is not produced by free-form generation. Instead, retrieval-only provides the highest coverage, while safe hybrid verification provides the highest accepted precision.

\begin{table}[t]
\caption{Compact batch results. Each cell reports coverage / AcceptBoth (\%).}
\label{tab:main-results}
\centering
\small
\begin{tabular}{@{}lcccc@{}}
\toprule
\textbf{Config} & \textbf{Label} & \textbf{Free} & \textbf{Retrieve} & \textbf{Hybrid} \\
\midrule
VL2-Base & 11 / 100 & 0 / 0 & 100 / 62 & 50 / 78 \\
VL2-FT   & 28 / 80  & 0 / 0 & 100 / 62 & 50 / 78 \\
Q2VL-Base& 0 / 0    & 0 / 0 & 100 / 62 & 50 / 78 \\
Q2VL-FT  & 0 / 0    & 0 / 0 & 100 / 62 & 50 / 78 \\
\bottomrule
\end{tabular}
\end{table}

In the main text, we foreground the key trend: free generation produces no accepted attributions, retriever-only provides full coverage, and hybrid verification improves accepted precision by abstaining.

Four findings emerge.

\paragraph{Open-ended generation fails under conservative museum conditions.}
All four model configurations produce 0\% coverage in free-generation mode. This shows that unrestricted VLM generation is not a reliable basis for catalogue attribution in in-gallery video, especially when the system is required to abstain instead of guessing.

\paragraph{Catalogue retrieval is the strongest high-coverage identification engine.}
Retriever-only mode achieves 100\% coverage and 62\% title--artist correctness among accepted outputs. This confirms that the finite museum catalogue is not just auxiliary context; it is the central source of attribution reliability.

\paragraph{Safe hybrid verification improves accepted precision by abstaining.}
Hybrid verification reduces coverage from 100\% to 50\%, but increases accepted title--artist correctness from 62\% to 78\%. This is the desired behaviour for a metadata insertion setting, where a false positive is more harmful than an abstention.

\paragraph{Fine-tuning is task-specific rather than universally beneficial.}
VideoLLaMA2 fine-tuning improves label-first coverage from 11\% to 28\%, although accepted precision drops from 100\% to 80\%. Qwen2-VL fine-tuning does not improve label-first or free-generation performance in the batch evaluator, and the safe hybrid results are identical for base and fine-tuned Qwen2-VL. The current gain therefore comes mainly from retrieval grounding and safe decision calibration, not from LoRA alone.

\subsection{Ablation Studies}
\label{sec:ablation}

We ablate three components using the expanded evaluation data.

\subsubsection{Effect of Fine-Tuning}

Fine-tuning has a measurable but limited effect. In label-first mode, VideoLLaMA2 fine-tuning increases coverage from 11\% to 28\% and raises overall title--artist correctness from 12\% to 25\%, but accepted precision drops from 100\% to 80\% because one accepted prediction is wrong. Qwen2-VL fine-tuning does not improve label-first or free-generation performance in the batch evaluator.

In retrieval-only and hybrid-verification modes, all model configurations obtain identical final results because the dominant decision signal is the visual catalogue retriever and the safe policy. This means the system contribution is architectural: catalogue grounding and abstention matter more than unrestricted fine-tuning.

\subsubsection{Identity Source and Decision Policy}

Table~\ref{tab:main-results} provides the identity-source and decision-policy ablation. The key comparison is not between catalogue variants alone, but between unrestricted generation, catalogue retrieval, and abstention-aware hybrid verification.

\subsubsection{Threshold Sensitivity}

The safe hybrid policy exposes a controllable precision--coverage trade-off. A high-confidence top-1 threshold of 0.88 with a margin requirement of 0.04 accepts strong matches such as \emph{The Entombment}, \emph{Cupid complaining to Venus}, \emph{Whistlejacket}, \emph{A Young Woman seated at a Virginal}, and \emph{Exhibition of a Rhinoceros at Venice}. It also abstains on several cases that retriever-only accepts, including some correct Red Boy variants with scores below the high-confidence threshold. This behaviour is intentional: the archive-facing mode is designed to reduce false positives even when this sacrifices coverage.

\begin{figure}[t]
    \centering
    \includegraphics[width=\linewidth]{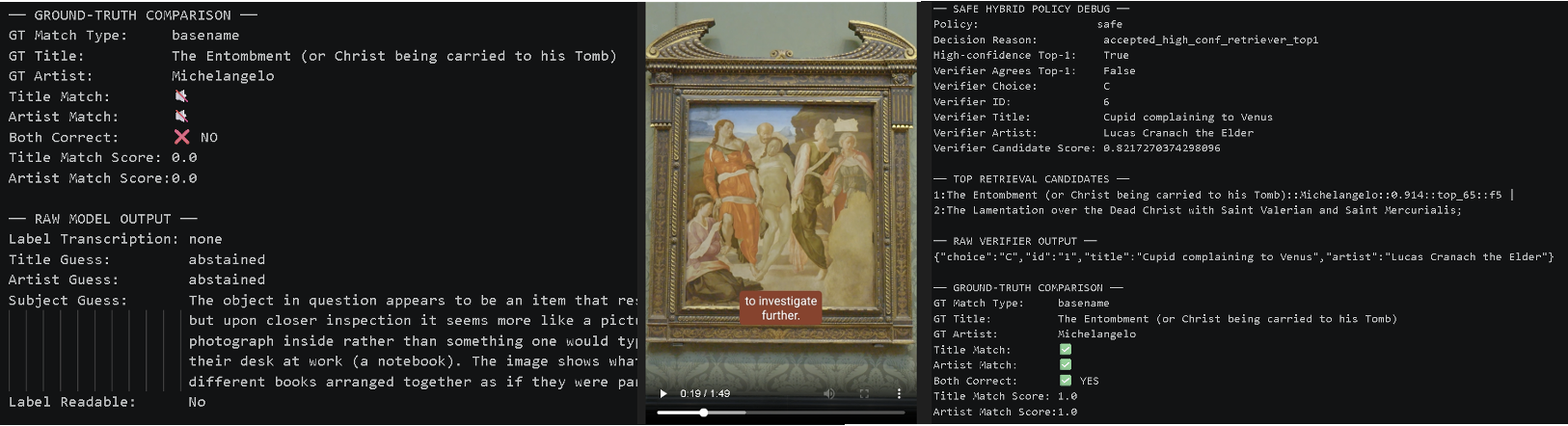}
    \caption{Examples of catalogue-grounded attribution on in-gallery video. The comparison illustrates the difference between free-form generation and catalogue-grounded retrieval/verification using the fine-tuned Qwen2-VL backend.}
\label{fig:hybrid-retrieval}
\end{figure}

\subsection{Qualitative Analysis}

Representative qualitative cases show the same pattern as the aggregate results. Catalogue-grounded modes recover clear matches such as \emph{The Entombment}, \emph{Cupid complaining to Venus}, \emph{Vermeer}, and \emph{Exhibition of a Rhinoceros at Venice}; \emph{The Hay Wain} illustrates a retrieval error against a visually related landscape candidate; and the Red Boy variants show that safe hybrid verification may abstain on otherwise correct but lower-confidence retrievals. These cases show how retrieval recovers many true matches while the safe hybrid mode remains deliberately conservative and still requires improved verifier calibration.

\subsection{System Constraints and Limitations}

\paragraph{Evaluation scale.}
We evaluate on 18 videos across four configurations and four modes, yielding 288 total system evaluations. The GT-labelled set remains modest because annotation requires institutional and curatorial validation, but the expanded evaluation is sufficient to reveal stable operating-mode differences.

\paragraph{Fine-tuning is aligned but not the primary source of hybrid gains.}
The latest setup trains both VideoLLaMA2.1 and Qwen2-VL as retrieval-conditioned verifiers on the same 672-record hybrid-verifier dataset. This improves experimental symmetry, but the final safe-hybrid results remain driven primarily by catalogue retrieval, thresholding, margin checks, and abstention calibration. We therefore avoid claiming that LoRA alone solves artwork attribution.

\paragraph{Training--inference alignment.}
Multimodal fine-tuning remains sensitive to the visual input pathway. The current setup reduces this risk by using precomputed crop-image verifier records for both training and evaluation, but broader investigation of format sensitivity across model families remains ongoing.

\paragraph{Precision--coverage trade-off.}
The safe hybrid policy increases accepted precision by abstaining, but it can reject true positives when retrieval confidence is below the high-confidence threshold. This is appropriate for conservative archive metadata insertion, but a discovery/search interface may prefer the retriever-only operating point.

\paragraph{Long, heterogeneous video.}
We process short clips rather than full-length assets to reduce cost and attention drift.


\section{\uppercase{Discussion}}

The experiments indicate that museum-video artwork attribution is better treated as a retrieval-and-decision problem than as unrestricted multimodal generation.

First, the finite catalogue provides a strong institutional prior. In a museum or gallery environment, the set of plausible identities is often much smaller than the open visual world. Ignoring this structure forces the model to solve a harder problem than the application actually requires.

Second, retrieval and generation have complementary rather than interchangeable roles. Retrieval is effective for grounding identity in canonical records, whereas generative models remain valuable for producing descriptions, scene summaries, semantic observations, and other metadata that cannot be represented as simple catalogue lookups.

Third, abstention should be treated as an explicit system capability rather than as a failure state. In a conservative archive workflow, a prediction can be represented as one of three operational outcomes:

\begin{equation}
\begin{aligned}
&\text{accept automatically},\quad \text{refer for review},\\
&\text{insufficient evidence}.
\end{aligned}
\end{equation}

This differs from conventional benchmark settings in which every sample is expected to receive a label. For institutional metadata systems, calibration and selective prediction may therefore matter as much as raw top-1 accuracy.

Finally, our results show that fine-tuning should be evaluated within the complete retrieval-and-decision architecture rather than in isolation. A lower training loss or increased willingness to answer does not necessarily improve safe downstream attribution. Fine-tuning that increases coverage while reducing accepted precision may be undesirable in a high-risk metadata insertion workflow.


\section{\uppercase{Conclusion}}

This work demonstrates that museum audiovisual archives can be approached as a practical and methodologically distinctive problem for computer vision in art and culture. By combining open video-language models, structured catalogue grounding, visual catalogue retrieval, and configurable abstention, the proposed framework converts in-gallery video into searchable, catalogue-compatible metadata while reducing the risk of unsupported attribution.

The results show that open-ended VLM generation is not reliable for this setting, while closed-set catalogue retrieval provides a strong high-coverage baseline. A safe hybrid decision policy further increases accepted precision by abstaining on uncertain cases, making it more appropriate for conservative archive metadata insertion.

More broadly, the study positions cultural heritage video as a valuable testbed for responsible multimodal learning: systems must support search and discovery, but also respect institutional constraints, rights-sensitive collections, and the curatorial consequences of false identification. Future work will extend the evaluation across larger and more diverse collections, integrate curator-in-the-loop validation, and investigate verifier calibration and retrieval confidence estimation for wider forms of visual-cultural analysis beyond artwork identification.


\section{\uppercase{Impact Statement}}

Catalogue-grounded multimodal attribution has the potential to unlock museum audiovisual archives by producing searchable, catalogue-linked metadata under local deployment constraints. Because misattribution can propagate harm, the system is designed to abstain explicitly when evidence is weak and to support curator-in-the-loop review. Responsible use requires governance, calibration, and monitoring to manage configuration errors, catalogue bias, retrieval bias, and misuse outside the heritage setting.


\section{\uppercase{Reproducibility}}

The relevant code will be released in due course. The backend abstraction enables reproduction with any compatible vision-language model. Training data is derived from publicly accessible museum catalogue records; evaluation videos are subject to institutional access agreements. The batch evaluation uses deterministic frame sampling, fixed retrieval settings, and explicit operating modes (\texttt{label\_first}, \texttt{free\_generation}, \texttt{retriever\_only}, \texttt{hybrid\_verify}) to support reproducibility across platforms.

\subsection{Demonstration}

An anonymised project demonstration can be provided as supplementary material for review. The public project page will be made available in the camera-ready version.


\bibliographystyle{apalike}
{\small
\bibliography{cagmula_scitepress}}

\end{document}